\documentstyle[prl,aps,twocolumn,epsf]{revtex}
\begin{document}
\draft
\title{Local functional models of critical correlations in thin-films}

\author{A.O.\ Parry, E.D.\ Macdonald and C.\ Rasc\'{o}n}
\address{Department of Mathematics,\\
Imperial College, London. SW7 2BZ}
\date{\today}
\maketitle

\begin{abstract}
Recent work on local functional theories of critical inhomogeneous fluids and Ising-like magnets has shown them to be a potentially exact, or near exact, description of universal finite-size effects associated with the excess free-energy and scaling of one-point functions in critical thin films. This approach is extended to predict the two-point correlation function $G$ in critical thin-films with symmetric surface fields in {\it{arbitrary}} dimension $d$. In $d\!=\!2$ we show there is {\it{exact}} agreement with the predictions of conformal invariance for the complete spectrum of correlation lengths $\xi^{(n)}$ as well as the detailed position dependence of the asymptotic decay of $G$. In $d\!=\!3$ and $d\!\geq\!4$ we present new numerical predictions for the universal finite-size correlation length and scaling functions determining the structure of $G$ across the thin-film. Highly accurate analytical closed form expressions for these universal properties are derived in arbitrary dimension.
\end{abstract}

Finite-size scaling theory predicts that the free-energy, order-parameter and correlation length of a critical thin-film Ising-like magnet (or of a fluid between parallel-plates at the bulk critical point) are characterised by universal amplitudes and scaling functions that depend only on dimensionality $d$ and the qualitative nature of of the boundary conditions \cite{Dg-F,Priv-F}. These have been the topic of widespread theoretical interest \cite{KrechCasimir} with approaches including sophisticated $\varepsilon$-expansion treatments as $d\!\rightarrow\!4$ \cite{Krech} and in particular studies that invoke conformal invariance in $d\!=\!2$ \cite{Cardy}. In a recent article Boran and Upton (BU) \cite{BU} have revisited the problem of critical thin-films using a local-functional model similar to one introduced earlier by Fisher and Upton \cite{Fsh-Up}. Specifically BU consider critical films of thickness $L$ and infinite area $A$ with both like sign $(++)$ and opposite sign $(+-)$ symmetry-breaking surface fields and obtain explicit expressions for both the universal Casimir amplitude $A_{ab}$ associated with the finite-size free-energy and universal scaling function $\Psi_{ab}(x/L)$ for the equilibrium magnetisation profile $m(x)$, for arbitrary dimension $d$. For the $(++)$ geometry in particular where there is less ambiguity in the choice of functional, the predictions of BU are in astonishing agreement with previous results obtained from the $\varepsilon$-expansion and conformal invariance. For example in $d\!=\!2$ the local functional result for $\Psi_{++}(x/L)$ is exact \cite{Burk-ErEn} whilst the numerical value obtained for $A_{++}$ is in near perfect agreement with the conformal result $A_{++}\!=\!- \pi / 48$ \cite{Conformal}. Whilst the local functional theory is phenomenological in nature, the agreement with microscopic theory is compelling and the model offers the exciting possibility of predicting (near) exact results for universal amplitudes and scaling functions in $d = 3$.

In the present work we extend the local functional theory to allow calculation of the two-point connected correlation function $G({\bf{r_{1}}}, {\bf{r_{2}}}) = \langle m({\bf{r_{1}}}) m({\bf{r_{2}}})\rangle - \langle m({\bf{r_{1}}}) \rangle \langle m({\bf{r_{2}}}) \rangle$ for the $(++)$ thin-film. First we discuss the $d\!=\!2$ Ising strip and compare the theory with exact results obtained from conformal invariance. This is a stringent test of the accuracy of the local functional theory and our analysis shows an extraordinary level of agreement which we find even more striking than the successes obtained by BU for $A_{++}$ and $\Psi_{++}(x/L)$ mentioned above. In particular we derive the precise spectrum of universal correlation lengths as well as the exact and highly non-trivial position dependence of the leading-order and next-to-leading order asymptotic decay of $G$. We are also able to obtain the exact Casimir amplitude $A_{++}$, improving on the estimate of BU. Secondly we present detailed predictions for the universal finite-size correlation length and scaling function $\Xi_{d}(x/L)$ characterising the asymptotic decay of $G$ for the $d\!=\!3$ and  $d\!\geq\!4$ thin-films. These features of $G$ have not been discussed in detail before and we are able to derive accurate analytic expressions for them valid in all dimensions.

To begin we recall some pertinent universal finite-size scaling properties of critical thin-films and the exact predictions for $G$ in $d\!=\!2$ arising from conformal invariance. Consider a $d$ dimensional Ising-like thin film with surface fields $h_{1}$, $h_{2}$ acting on the spins in the surface planes $x\!=\!0$ and  $x\!=\!L$. These fields mimic the preferential local adsorption of the confining walls on a fluid in a parallel-plate geometry. We define the reduced excess free-energy $f^{\times}(L)\!\equiv\!(F - A L f_{b})/ A k_{B} T$ by subtracting from the total free-energy $F$ the appropriate contribution from the bulk free-energy density $f_{b}$. By further subtracting from  $f^{\times}(L)$ the semi-infinite (reduced) surface tensions of the two-independent surfaces we obtain the desired finite-size contribution to the free-energy, $\Delta f^{\times}(l)$. Finite size scaling theory predicts that exactly at the bulk critical point, for asymptotically large widths $L$ (and dimension $d\!<\!4$), this can be written $\Delta f^{\times}(L)\!=\!A_{ab}/L^{d - 1} + \dots$ where the Casimir amplitude $A_{ab}$ depends only on whether the surface fields have like $(++)$ or opposite $(+-)$ sign (or are zero). Hereafter we restrict attention to the $(++)$ geometry where the results of the local functional theory are most striking. A second universal quantity is the normalised scaling function $\Psi_{++}(x/L)$ describing the magnetisation profile $m(x)$ across the thin-film. For a semi-infinite system the asymptotic large distance decay of the critical magnetisation profile is $m(x)\!\simeq\!C x^{- \beta / \nu}$ where $C$ is a non-universal amplitude and $\beta$, $\nu$ are standard bulk critical exponents for the magnetisation and correlation length respectively. In the $(++)$ critical thin-film finite-size scaling \cite{Dg-F} predicts that for large $x$, $L$ with $x/L$ arbitrary the profile can be written $m(x)\!=\!m(L/2) \Psi_{++}(x/L)$ where the mid-point magnetisation scales as $m(L/2)\!\propto\!L^{- \beta / \nu}$ and $\Psi_{++}(x/L)$ is normalised such that $\Psi_{++}(1/2)\!=\!1$. Finally, finite-size scaling theory \cite{Priv-F} also predicts that the asymptotic decay of correlations along the critical thin-film is described by a (true) correlation length $\xi^{(0)}$, with $L/ \xi^{(0)}$ a universal critical amplitude ratio.

It is well known that in $d\!=\!2$ these predictions are beautifully confirmed by conformal invariance \cite{Cardy} and are consistent with results obtained from exactly solvable models. In particular by using the standard logarithmic function to conformally map the semi-infinite plane to the finite-size strip the magnetisation follows \cite{Burk-ErEn} as $m(x)\!=\!C (\frac{L}{\pi} \sin \frac{\pi x}{L})^{-1/8}$ implying $\Psi_{++}(z)\!=\!(\sin \pi x)^{-1/8}$. Applying the same mapping to the semi-infinite two-point function calculated by Burkhardt and Xue \cite{Burk-Xue} we obtain
\begin{eqnarray}
G^{++}(x_{1}, x_{2}; y) & = & \frac{C^{2}}{\sqrt{2}} \left ( \frac{L}{\pi} \right )^{-1/4} \left ( \sin \frac{\pi x_{1}}{L} \sin \frac{\pi x_{2}}{L} \right )^{-1/8} \times \nonumber\\
& & \hspace{1.4cm} \left [ \sqrt{u + u^{-1}} - \sqrt{2} \, \right ]
\end{eqnarray}
\vspace{-0.5cm}
\begin{eqnarray}
\hspace{-0.8cm} \hbox{where} \hspace{0.5cm} u = \left [ \frac{\cosh \pi y / L - \cos \pi (x_{2} + x_{1})/L}{\cosh \pi y / L - \cos \pi (x_{2} - x_{1})/L} \right ]^{1/4}\nonumber
\end{eqnarray}
and $y$ denotes the scalar distance between the spins measured along the strip. From the asymptotic $y/L\!\rightarrow\!\infty$ expansion of (2) it follows that the true correlation length in the $(++)$ strip is $\xi^{(0)}\!=\!L/ 2 \pi$ identifying the universal critical amplitude ratio predicted by finite-size scaling theory. More generally, this expansion identifies the spectrum of correlation lengths $\xi^{(n)}$ determining the decay of the exponential terms as
\begin{equation}
\xi^{(n)} = \frac{L}{(n + 2) \pi}\; ; \hspace{0.25cm} \hbox{$n = 0$, $1$, $2$, $3$, $\dots$}.
\end{equation}
Thus the explicit conformal invariance prediction for the leading-order and next-to-leading-order asymptotic $y / L\!\rightarrow\!\infty$ decay of $G^{++}$ is
\begin{eqnarray}
G^{++}(x_{1}, x_{2}; y) = \frac{C^{2}}{4} \left ( \frac{L}{\pi} \right )^{-1/4} \left ( \sin \frac{\pi x_{1}}{L} \sin \frac{\pi x_{2}}{L} \right )^{15/8} \times \nonumber
\end{eqnarray}
\vspace{-0.5cm}
\begin{eqnarray}
\hspace{0.5cm} e^{-2 \pi y /L} \left \{ 1 + 4 \cos \frac{\pi x_{1}}{L} \cos \frac{\pi x_{2}}{L} e^{-\pi y /L} + \dots \right \}
\end{eqnarray}
which shows a detailed position dependence across the strip. We wish to compare these exact predictions with the results of the BU local functional theory.

The desirable scaling properties of quite general local functionals of the magnetisation have been considered by Fisher and Upton \cite{Fsh-Up} but in application to the $(++)$ thin-film (where there is no zero of the equilibrium profile) the appropriate model is simplicity itself. BU assume translational invariance of the trial profiles along the thin-film and base their analysis on the excess (reduced) free-energy functional (ignoring boundary terms)
\begin{equation}
{\mathcal{F}}[m] = \int_{0}^{L} dx \left [ \frac{K}{2} m^{- \eta \nu / \beta} \dot{m}^{2} + \frac{u}{\delta + 1} m^{\delta + 1}\right ]\nonumber
\end{equation}
where $\dot{m}\!\equiv\!d m / d x$, $\eta$ and $\delta$ are standard bulk exponents and $K$, $u$ are parameters that must be regarded as inputs into the theory, although their values do not influence the determination of the scaling function $\Psi_{++}(z)$ nor the structural properties of $G$. The surface fields enter through boundary terms but these play no role in determining the universal quantities of interest and may be replaced by fixed boundary conditions $m(0)\!=\!m(L)\!=\!\infty$ as appropriate to the scaling limit. Specialising to $d\!=\!2$ and using the bulk critical exponents $\beta\!=\!1/8$, $\nu\!=\!1$, $\eta\!=\!1/4$ and $\delta\!=\!15$ the Euler-Lagrange equation for the equilibrium profile is
\begin{equation}
K m^{-2} \ddot{m} - K m^{-3} \dot{m}^{2} = u m^{15}
\end{equation}
which has a first integral
\begin{equation}
\frac{K m^{-2} \dot{m}^{2}}{2} = \frac{u m^{16}}{16} - \Delta p
\end{equation}
where the constant of integration $\Delta p\!=\!\partial f^{\times} / \partial L\!=\!- A_{++}L^{-2}$. From (5) it is straight forward to derive \cite{BU}
\begin{equation}
m(x) = \left ( \frac{K}{8u}\right )^{1/16} \left (\frac{L}{\pi} \sin \frac{\pi x}{L}\right )^{- 1/8}
\end{equation}
in exact accordance with conformal invariance. From the profile it follows that $\Delta p\!=\!\pi^{2} K / 128 L^{2}$ implying $A_{++}\!=\!- \pi^{2} K / 128$. Using scaling arguments BU relate $K$ to known bulk critical exponent and amplitude ratios and yield a numerical value for $A_{++}$ extremely close to the exact result $A_{++}\!=\!- \pi/48$ \cite{Conformal} and with an error that may even be attributed to uncertainties in the values of bulk critical amplitude ratios used to determine $K$. We shall determine $K$ another way and show how the exact result for $A_{++}$ may be obtained within the local functional theory.

With these preliminaries over we can now turn to the main part of our work and describe the calculation of $G^{++}$ within the local functional theory. As it stands the BU functional (4) is not capable of yielding the full correlation function because it assumes translational invariance of the magnetisation along the strip. However it is reasonable to argue from isotropy that the full functional is given by
\begin{equation}
{\mathcal{F}}[m] = \frac{1}{A} \int \int dy \, dx \left [ \frac{K}{2} m^{-2} (\nabla m)^{2} + \frac{u}{16} m^{16}\right ]
\end{equation}
and recall that we have adopted fixed boundary conditions as appropriate to the scaling limit. Minimisation of (8) leads to the extended Euler-Lagrange equation
\begin{equation}
K m^{-2} \nabla^{2} m - K m^{-3}(\nabla m)^{2} = u m^{15}
\end{equation}
which of course reduces to (5) as soon as we impose translational invariance along the strip. We mention here that the solution to (9) for the magnetisation profile interior to a disk of radius R with fixed up-spins at the boundary is $m(r)\!=\!C(\frac{R}{2}(1 - r^{2}/ R^{2}))^{-1/8}$ which is identical to the exact conformal invariance result \cite{Burk-ErEn}. The same is also true in all higher dimensions $d$ if we adopt the appropriate values for the bulk critical exponents in (9).

Within the framework of density-functional theory the two-point correlation function follows from solution of the Ornstein-Zernike equation \cite{Evans}
\begin{equation}
\int {\bf{dr_{3}}} \, \frac{\delta^{2} {\mathcal{F}}[m]}{\delta m({\bf{r_{1}}}) \, \delta m({\bf{r_{3}}})} \, G({\bf{r_{2}}}, {\bf{r_{3}}}) = \delta({\bf{r_{1}}} - {\bf{r_{2}}}).
\end{equation}
For our functional this reduces to the differential equation
\begin{eqnarray}
\left[ \frac{135}{8} u m^{14} + 2 m^{-2} \Delta p - K \frac{\partial}{\partial x_{1}} m^{-2} \frac{\partial}{\partial x_{1}}
\right. \hspace{1.3cm}\nonumber \\
\left. - K m^{-2} \frac{\partial^{2}}{\partial y^{2}} \right] G^{++}(x_{1}, x_{2}; y) = \delta(x_{1} - x_{2}) \delta(y)
\end{eqnarray}
which we seek to solve using the spectral expansion
\begin{equation}
G^{++}(x_{1}, x_{2}; y) = \sum_{n = 0}^{\infty} \psi_{n}(x_{1}) \psi_{n}^{*}(x_{2}) e^{- E_{n} y}
\end{equation}
with energy levels $E_{n}$ equivalent to the inverse correlation lengths $1/\xi^{(n)}$ discussed earlier. Orthonormality implies
\begin{equation}
2 K \int_{0}^{L} dx \psi_{n}^{*}(x) m(x)^{-2} \psi_{m}(x) = \frac{\delta_{n m}}{E_{n}}
\end{equation}
which is needed to determine the correct normalisation factors of the eigenstates. The spectrum is thus obtained through solution of the eigenvalue problem
\begin{eqnarray}
\left[ \frac{135}{8} u m^{16} + 2 \Delta p - K m^{2} \frac{\partial}{\partial x} m^{-2} \frac{\partial}{\partial x} \right] \psi_{n}(x) \hspace{0.25cm}\nonumber\\
= K E_{n}^{2} \psi_{n}(x)
\end{eqnarray}
subject to the boundary condition $\psi_{n}(0)\!=\!\psi_{n}(L)\!=\!0$. Substituting for $m(x)$ and the value of $\Delta p$ quoted earlier (without having to fix $K$) this reduces to
\begin{eqnarray}
- \psi_{n}''(x) - \frac{\pi}{4 L} \left ( \cot \frac{\pi x}{L} \right ) \psi_{n}'(x) + \hspace{0.2cm}& & \nonumber \\
\left [ \frac{135 \pi^{2}}{64 L^{2}} \left (\sin \frac{\pi x}{L} \right )^{-2} + \frac{\pi^{2}}{64 L^{2}} \right ] \psi_{n}(x) & = & E_{n}^{2} \psi_{n}(x).
\end{eqnarray}
which can be solved using standard methods \cite{M&P}. For the eigenvalue spectrum we find
\begin{equation}
E_{n} = \frac{(n + 2) \pi}{L}\;; \hspace{0.25cm}\hbox{$n = 0$, $1$, $2$, $3$, $\dots$}
\end{equation}
whilst the eigenvectors can be written as $\psi_{n}(x)\!=\!N_{n} (\sin \frac{\pi x}{L})^{15/8} \phi_{n}(x)$ with
\begin{equation}
\phi_{n}(x) = F \left [- \frac{n}{2}, 2 + \frac{n}{2}, \frac{1}{2}; \cos^{2} \frac{\pi x}{L} \right ]
\end{equation}
for the even states and
\begin{equation}
\phi_{n}(x) = \cos \frac{\pi x}{L} F \left [\frac{1}{2} - \frac{n}{2}, \frac{5}{2} + \frac{n}{2}, \frac{3}{2}; \cos^{2} \frac{\pi x}{L} \right ]
\end{equation}
for the odd states. Here $F[\alpha, \beta, \gamma; z]$ denotes the usual hypergeometric function and $N_{n}$ is an appropriate normalisation factor which can be found using (13). The eigenvalue spectrum (16) is identical to the result obtained from conformal invariance (2) and exactly identifies all the universal critical amplitude ratios for the correlation lengths. From the above the asymptotic decay of the correlation function within the local functional theory is given by
\begin{eqnarray}
G^{++}(x_{1}, x_{2}; y) = \frac{C^{2}}{4} \frac{8}{3 K \pi} \left (\sin \frac{\pi x_{1}}{L} \sin \frac{\pi x_{2}}{L} \right )^{15/8} \times
\end{eqnarray}
\vspace{-0.75cm}
\begin{eqnarray}
e^{-2 \pi y/ L} \left \{ 1 + \left (\frac{N_{1}}{N_{0}} \right )^{2} \cos \frac{\pi x_{1}}{L} \cos \frac{\pi x_{2}}{L} e^{- \pi y/ L} + \dots \right \}\nonumber
\end{eqnarray}
where $C\!=\!(K/8 u)^{1/16}$ is the non-universal constant for the magnetisation while from (13) we find $N_{1}/N_{0}\!=\!2$ in precise agreement with the conformal result (3)! Moreover it is clear that the parameter $K$ simply sets the overall scale of the correlation function and that in $d\!=\!2$ the appropriate value is $K\!=\!3 \pi / 8$. Utilising this identification we are led to the prediction $A_{++}\!=\!-\pi/48$ which is identical to the exact conformal invariance result \cite{Burk-Xue}. We comment here that only small discrepancies with the conformal result are found at higher orders which are surely of no practical importance. It is also possible to calculate the moments $G^{++}_{2n}(x_{1}, x_{2})\!\equiv\!\int dy\, y^{2n} G^{++}(x_{1}, x_{2}; y)$ using the local functional theory and compare these with the expressions obtained from conformal invariance. Consider for example the zeroth moment $G_{0}(x_{1}, x_{2})$ from which we can construct a universal two-point scaling-function $\Lambda^{++}_{0}(x_{1}/L, x_{2}/L)\!=\!G^{++}_{0}(x_{1}, x_{2})/G^{++}_{0}(L/2, L/2)$. We omit the details of this calculation and simply quote our final result calculated within the BU functional \cite{M&P}
\begin{eqnarray}
\Lambda^{++}_{0} \left (\frac{x_{1}}{L}, \frac{x_{2}}{L} \right ) = \left (\sin \frac{\pi x_{1}}{L} \sin \frac{\pi x_{2}}{L} \right)^{-1/8} \times \hspace{1cm}\nonumber\\
\left [ \frac{\pi x_{1}}{L} \cot \frac{\pi x_{1}}{L}\right ] \left [ \pi \left ( \frac{x_{2}}{L} - 1 \right ) \cot \frac{\pi x_{2}}{L} - 1 \right ]
\end{eqnarray}
which turns out to be identical to the conformally invariant result obtained by (numerically) integrating (2) provided that $x_{1}\!=\!x_{2}$. There are small discrepancies between the expressions if $x_{1}\!\neq\!x_{2}$ but even here the local functional theory is an excellent qualitative description of the structure of correlations across the strip. Nevertheless this implies that the exact conformal result for $G^{++}_{0}(x_{1}, x_{2})$ does not satisfy the correlation function algebra necessitated by all local functional theories \cite{P&S}. Given that it is therefore {\it{not}} possible to construct a truly exact local functional theory consistent with conformal invariance predictions for $G^{++}(x_{1}, x_{2}; y)$, the level of agreement achieved by the theory is all the more striking.

Next we turn our attention to the $d$ dimensional thin-film and focus on the decay of $G^{++}(x_{1}, x_{2}; {\bf{y}})$ for asymptotic parallel displacements $\vert {\bf{y}} \vert\!\rightarrow\!\infty$. Within the local functional theory this behaves as
\begin{equation}
G^{++}(x_{1}, x_{2}, {\bf{y}}) \sim C^{2} \Xi_{d}\left(\frac{x_{1}}{L}\right) \Xi_{d}\left(\frac{x_{2}}{L}\right) \frac{e^{- \vert \, {\bf{y}} \vert / \xi^{(0)}}}{\vert \, {\bf{y}} \vert^{(d - 2)/2}}
\end{equation}
and recall that we anticipate that $L/ \xi^{(0)}$ is universal. Here we have introduced a universal scaling function $\Xi_{d}(x/L)$ describing the position dependence across the thin-film which we normalise (analogous to $\Psi_{++}$) such that $\Xi_{d}(1/2)\!=\!1$. Both $\Xi_{d}(x/L)$ and $\xi^{(0)}$ are determined by an eigenvalue equation analogous to (15) which depends only on $d$, $\beta / \nu$ and $\Psi_{++}(x/L)$. No information concerning $K$ or $u$ is required. We omit details of the calculation which are obtained assuming the same numerical values for the critical exponents $\beta\!=\!0.328$ and $\nu\!=\!0.632$ in $d\!=\!3$ as used by BU and the standard mean-field results $\beta\!=\!\nu\!=\!1/2$ for $d\!\geq\!4$. For the universal correlation length critical amplitude we obtain $L/ \xi^{(0)}\!\simeq\!7.34$ and $L/ \xi^{(0)}\!\simeq\!8.40$ in $d\!=\!3$ and $d\!\geq\!4$ respectively and recall $L/ \xi^{(0)}\!=\!2 \pi$ in $d\!=\!2$. The numerical results for the scaling functions $\Xi_{d}(x/L)$ are quite remarkable and are shown in Fig. 1. To extremely high precision (see inset) and for all practical purposes these functions are completely indistinguishable from the analytic expression
\begin{equation}
\widetilde{\Xi}_{d}(x/L) = \left ( \sin \frac{\pi x}{L} \right )^{d^{*} - \beta/ \nu}
\end{equation}
where $d^{*}\!=\!\min \{ d , 4 \}$. Notice this is exact in $d\!=\!2$ and  embodies the correct short-distance expansion behaviour close to each wall $\forall d$. Using $\widetilde{\Xi}_{d}(x/L)$ we can derive an extremely accurate expression for the ground state eigenvalue and hence obtain for the universal ratio
\begin{equation}
\frac{L}{\xi^{(0)}} = \sqrt{2 d^{*} \left (2 + \frac{\beta}{\nu} \right ) J_{++}(1)^{2} + \pi^{2} \left (d^{*} - \frac{\beta}{\nu} \right ) }
\end{equation}
where $J_{++}(1)\!=\!\int_{0}^{1} du\, (1 - u^{d^{*}})^{-1/2}$ is the elliptic function introduced by BU. This is exact in $d\!=\!2$ and within $0.3 \%$ of the numerical values quoted above for $d\!=\!3$ and $d\!\geq\!4$. For all intents and purposes this may be regarded as the exact analytical local functional prediction for the universal critical amplitude ratio.

In summary we have established that the BU local functional model is an excellent description of correlations in the $d\!=\!2$ critical $++$ thin-film and used the theory to make predictions for the universal structure of the correlation function in higher dimensional systems.

EDM and CR acknowledge support from the EPSRC and EC (ERBFM-BICT983229) respectively.
\vspace{-0.5cm}
\begin{figure}[h]
\epsfxsize=0.5\textwidth
\centerline{\epsffile{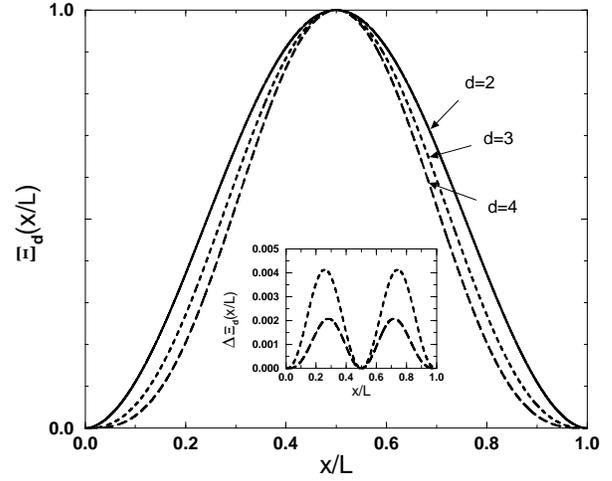}}
\vspace{-0.25cm}
\caption{Comparison of the universal functions $\Xi_{d}(x/L)$ in the $d\!=\!2$, $3$ and $4$ $(++)$ thin-films. Inset is shown the discrepancy $\Delta \Xi_{d}(x/L)\!=\!\Xi_{d}(x/L) - \widetilde{\Xi}_{d}(x/L)$ from the closed form approximation (22) which is not visible on the scale of the main figure.}
\end{figure}

\end{document}